\documentclass{PoS}

\title{Searches for SUSY in events with third-generation particles at CMS}

\ShortTitle{Searches for SUSY in events with third-generation particles at CMS}

\author{\speaker{Altan Cakir}\\
       {\rm On behalf of the CMS Collaboration}\\
       Deutsches Elektronen-Synchrotron (DESY)\\
       E-mail: \email{cakir@cern.ch}}

\abstract{Results of searches for SUSY production at CMS in events with third-generation signatures are presented. Along with missing energy, the final states may include hadronic jets with or without b-quark tag, light leptons, and tau leptons. These features serve both to distinguish standard-model components, and sensitivity to those SUSY models that lead to final states rich of heavy-flavored particles.}

\FullConference{36th International Conference on High Energy Physics\\
		 4-11 July 2012\\
		 Melbourne, Australia}

\begin{document}

\section{Introduction}

Supersymmetry (SUSY) is an extension of the Standard Model (SM) that naturally resolves the hierarchy problem~\cite{martin}. Within SUSY electroweak parameters are related to SUSY partners as shown in equation~ \ref{eq:mssm}. If the superpartners are too heavy, the contributions to the MSSM spectrum, for instance, must be tuned against each other to achieve electroweak symmetry breaking at the observed energy scale. 

\begin{equation}
\label{eq:mssm}
-M^{2}_Z/2 = |\mu^{2}| + m^{2}_{H^{2}_{u}}
\end{equation}

The masses of the superpartners with large couplings to the Higgs, i.e the stop and gluino masses, which correct $m^{2}_{H^{2}_{u}}$ must not be too far above the weak scale. In addition, the higgsinos should not be too heavy because their mass is controlled by the $\mu$ parameter. This has interesting implications if we consider the size of SUSY pair-production cross sections at 7 and 8 TeV center-of-mass energy.  These considerations suggest that the channel to search for SUSY involves the third generation of SUSY particles and their decays to top and bottom quarks and tau leptons. In the following sections several searches for SUSY in events with third-generation particles are discussed. In section 2, the final state with containing a single lepton, b-jets and missing tranverse momentum at $\sqrt{s}=7$ TeV is presented with two different data-driven estimation methods~\cite{singlelb}. In section 3, an analysis with one or more hadronically decaying $\tau$-leptons, highly-energetic jets and large momentum imbalance is discussed~\cite{tau}. In section 4, a search for hadronic final states with a b-tagged jet and missing transverse energy at $\sqrt{s}=7$ TeV is discussed~\cite{had}. In section 5, the preliminary results for a SUSY search in events with same-sign dileptons and b-tagged jets at $\sqrt{s}=8$ TeV are shown, together with the interpretations of simplified models~\cite{ss}. All analyses are performed using data collected with the Compact Muon Solenoid (CMS ) detector~\cite{Ball:2007zza} in proton-proton collisions at center-of-mass energies of 7 TeV and 8 TeV, corresponding integrated luminosities of $4.98 fb^{-1}$ and $3.96fb^{-1}$ respectively.

\section{Search for supersymmetry in events with a single lepton and b-tagged jets at $\sqrt{s}$= $7$ TeV}

The search is performed in two different data-based estimation methods. The signal regions are defined using the scalar sum of the jet transverse momenta ($H_T$), the missing transverse energy (MET), and the b-jet multiplicity. In the first approach, a factorization ansatz based on the variables $H_T$ and $Y_{MET}$ = MET/$\sqrt{H_T}$ is employed. In the second approach, templates for the MET spectra in W+jets and tt production are extracted from an inclusive single-lepton sample by a simultaneous fit to the 0, 1, and $\ge$ 2b-jet sub-samples. 

\begin{figure}[t]
\centering
\includegraphics[height=5cm]{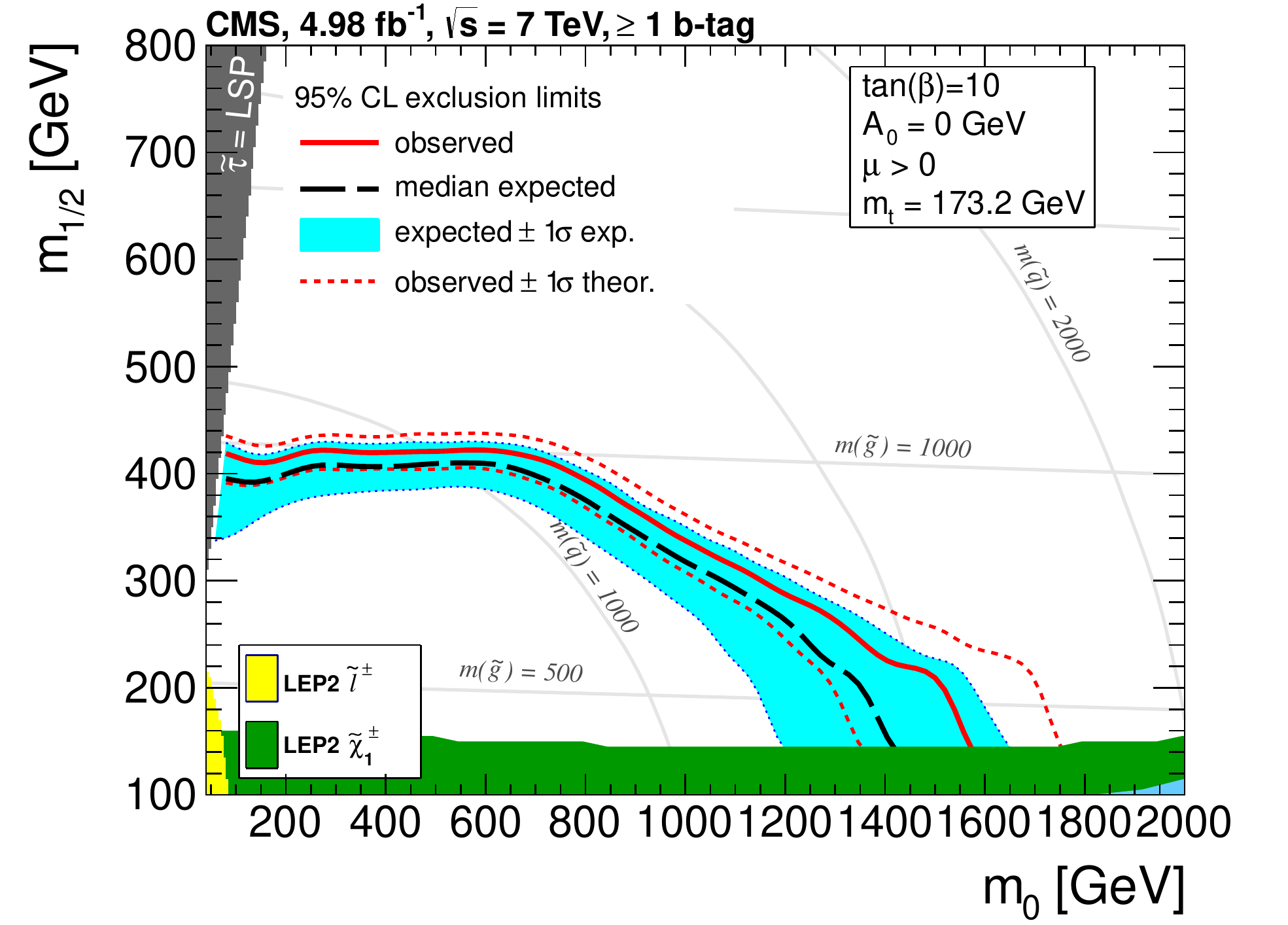}
\includegraphics[height=5.1cm]{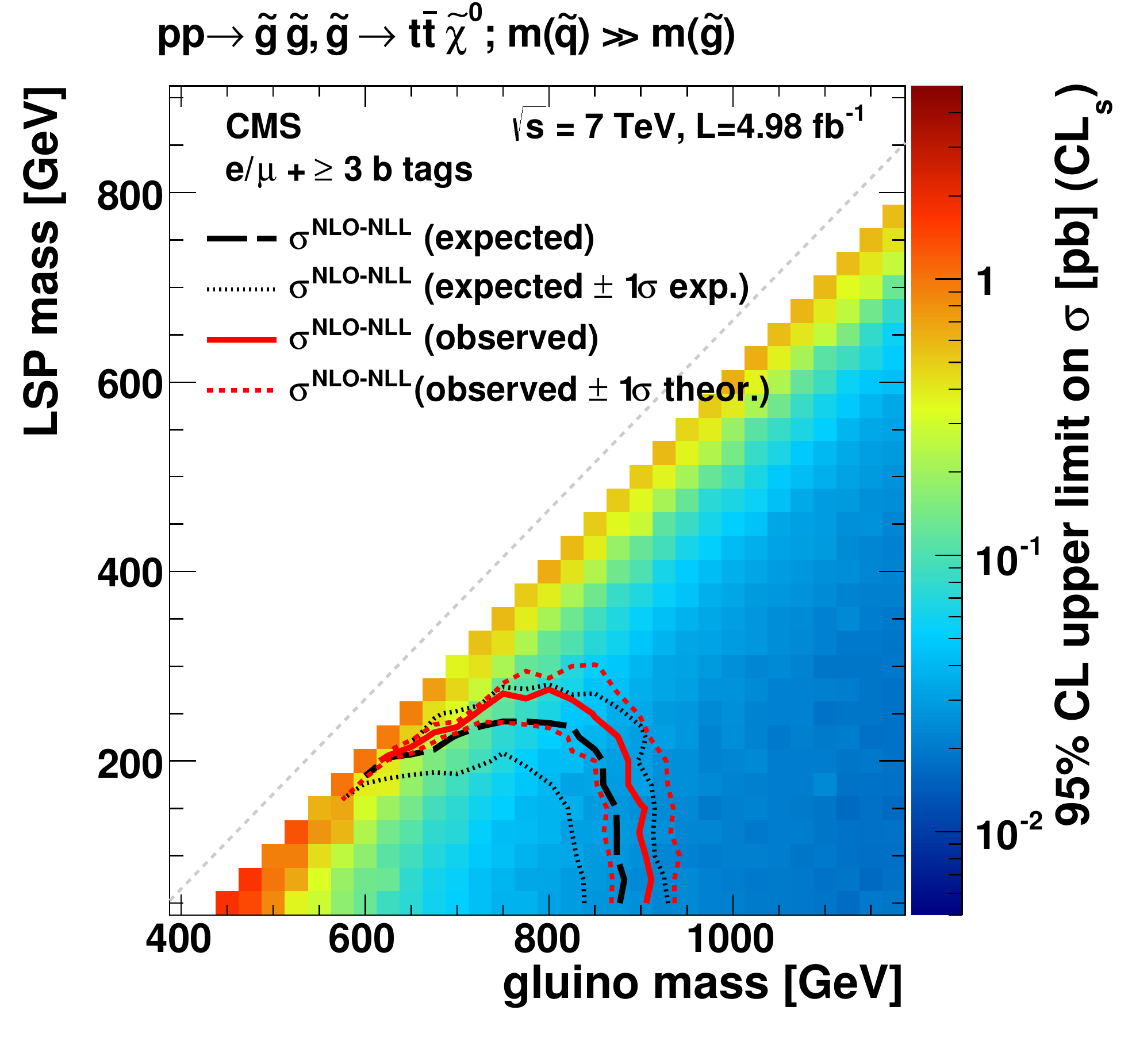}
\caption{The $95\%$ CL limit using the CLs technique for the CMSSM model with tan$\beta$=$10$, $A_{0}$=$0$ GeV and $\mu>0$ for the factorization method requiring at least one b tag (left) and the simplified model for three or more b tags (right). The area below the solid lines are excluded. }
\label{fig:first}
\end{figure}

The factorization method is based on the observation that the distribution of events is nearly uncorrelated with respect to the variables $H_T$ and $Y_{MET}$. This can be used to estimate the background in the signal region where the dominant background is $t\bar{t}$. The method uses control regions with low $H_T$ and/or $Y_{MET}$ called A, B and C. The signal region D is defined as $H_{T}>800$, $Y_{MET}>5.5$ $\sqrt{GeV}$ and for $\ge$ 1 b-tag and $H_{T}>600$, $Y_{MET}>6.5$ $\sqrt{GeV}$ for $\ge$ 3b-tags. The number of events expected in signal region is estimated from the three control regions by $\hat{N} = \kappa N_B N_C/N_A$. The factor kappa accounts for small deviation from the factorization approach and is determined by simulation. No deviation from the SM has been found and upper limits have been set on production cross-sections in the cMSSM model, and in a simplified model, where a gluino decays to $t\bar{t}$ and neutralino, shown in Fig.~\ref{fig:first}. 

\begin{figure}[b]
\centering
\includegraphics[height=5.1cm]{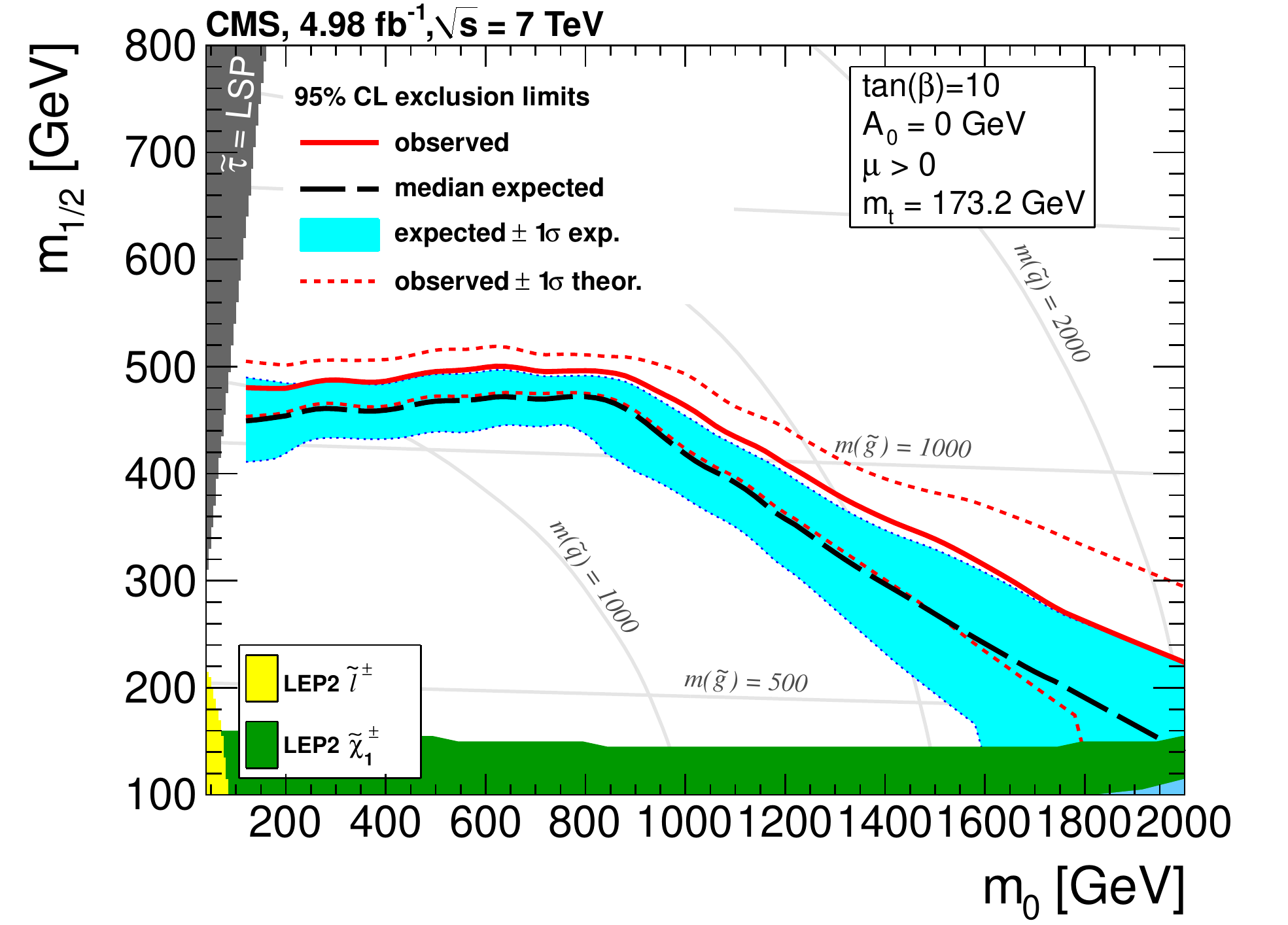}
\includegraphics[height=5.1cm]{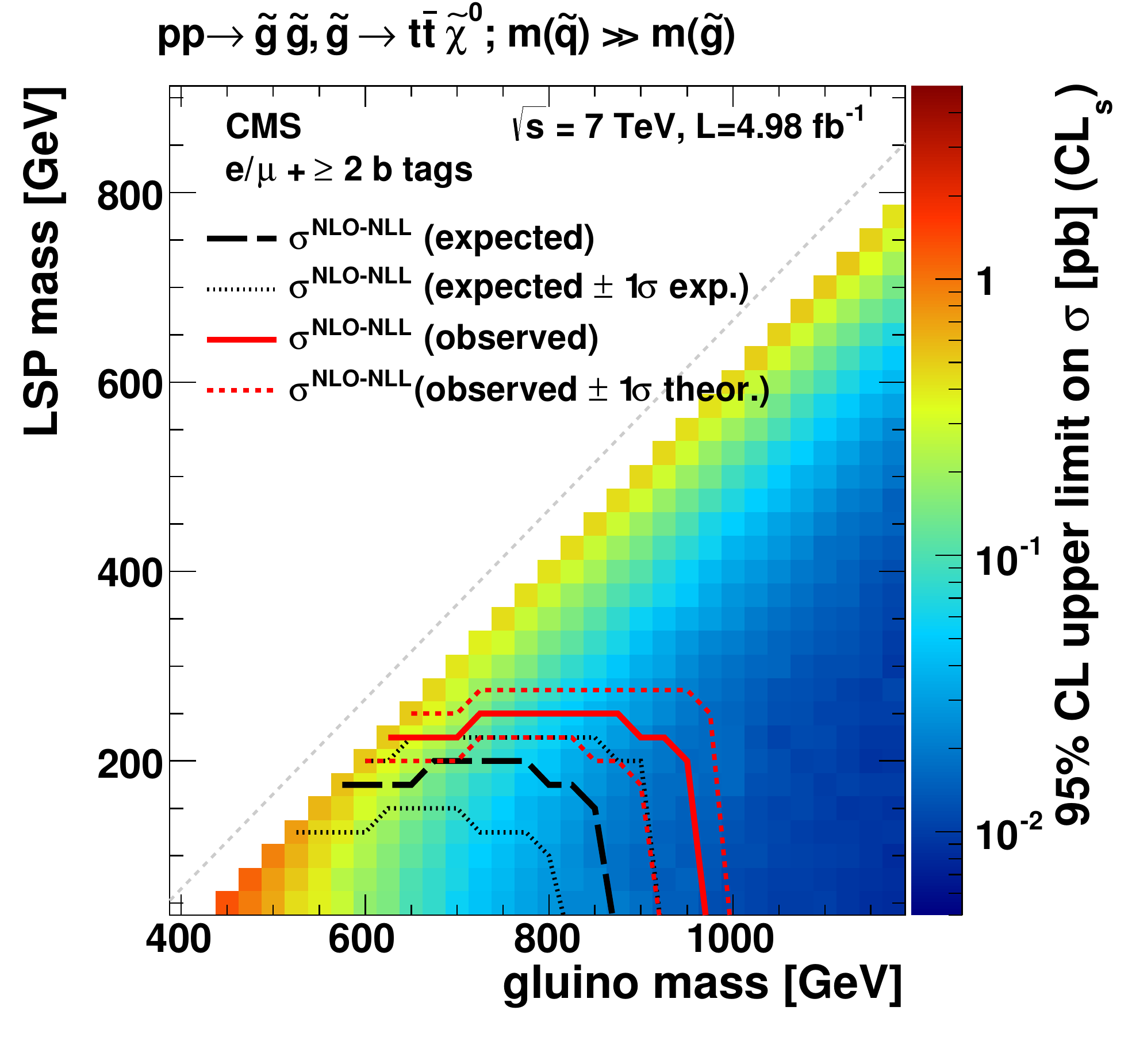}
\caption{The $95\%$ CL limit using the CLs technique for the CMSSM model with tan$\beta$=$10$, $A_{0}$=$0$ GeV and $\mu>0$ for the MET template method using the multichannel approach (left) and the simplified model requiring at least two b tags (right). The area below the solid lines are excluded.}
\label{fig:second}
\end{figure}

The MET template method is used to describe the rapidly falling MET distributions for the two main SM background ($t\bar{t}$ and W+jets). These distributions are different for $t\bar{t}$ and W+jets events, with details of the shape depending on the production process, i.e the polarization of the W, which is at the origin of the charged lepton. Therefore, a model describing the true MET distribution is convoluted with independently measured templates describing the detector response and fitted to data. A normalization region at high $H_T$ is then used to predict the background yields in signal regions at high $H_{T}$ and high MET. For the MET template method, the CMSSM limits are set in a multichannel approach using the 0 b-tag, 1 b-tag, and$\ge$2 b-tag bins, while for the factorization method at least one b tag is required. No excess has been observed and the results have been used to set 95\% CL exclusion limits for CMSSM and the simplified model, given in Fig.\ref{fig:second}. 

\section{Search for SUSY in events with $\tau$ leptons in the presence of multijets and large momentum imbalance at $\sqrt{s}$= $7$ TeV}

The dominant SM background contributions in this analysis are $t\bar{t}$, W and Z production with associated jets, where either real $\tau$ or jets misidentified as $\tau$. In addition, QCD multijet events contribute to the background when a badly mismeasured jet gives rise to large MET and jets are misidentified as $\tau$ leptons. Different methods have been used for single and di-$\tau$ final states. For the single $\tau$, the main background consist of events containing a real $\tau$ and events where a jet is misidentified as a $\tau$. In the $\tau\tau$ final state, the main contribution comes from one or more jets being misidentified as a $\tau$ for different background sources. 

\begin{figure}[hbt]
\centering
\includegraphics[height=6.5cm]{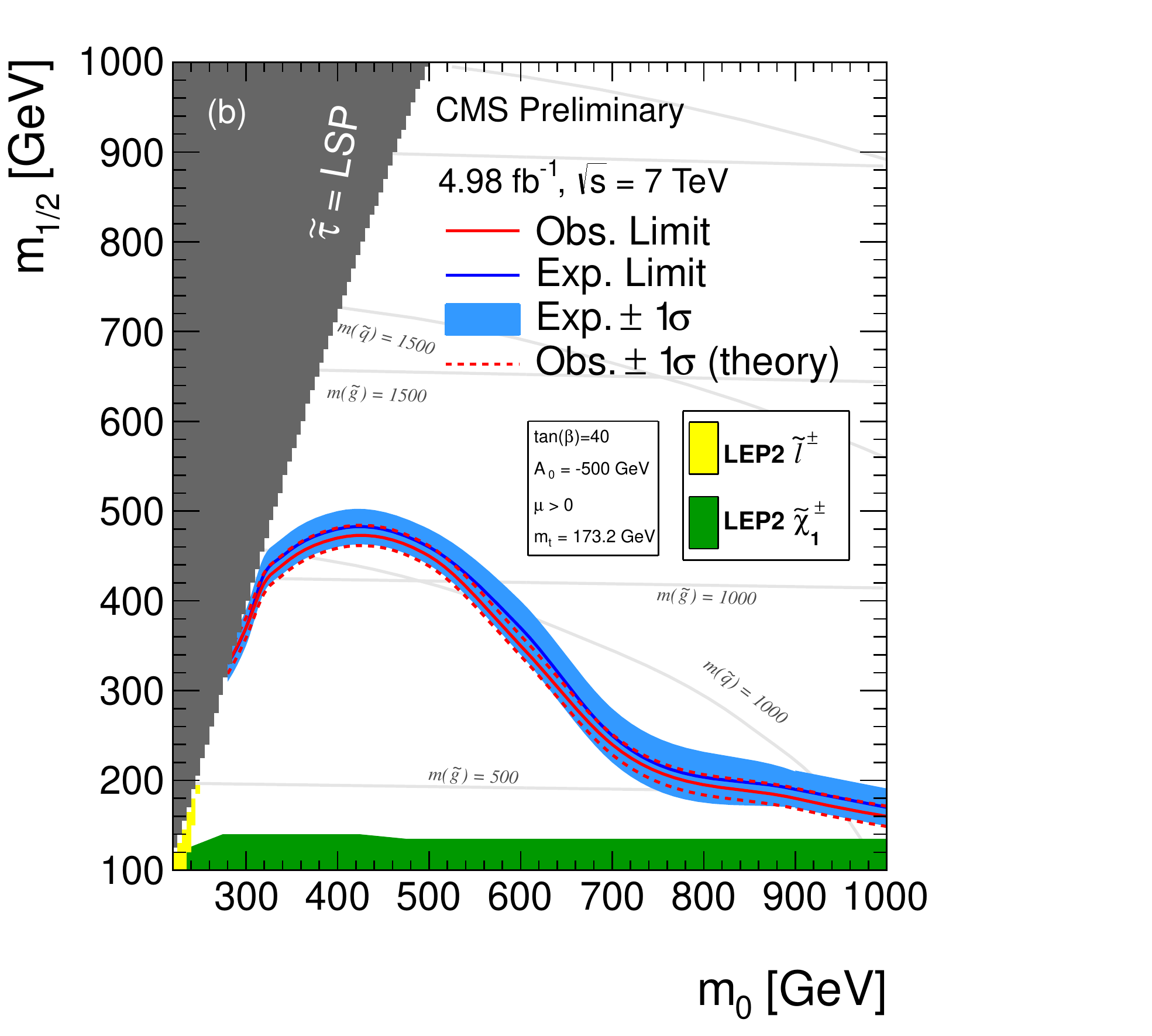}
\includegraphics[height=6.5cm]{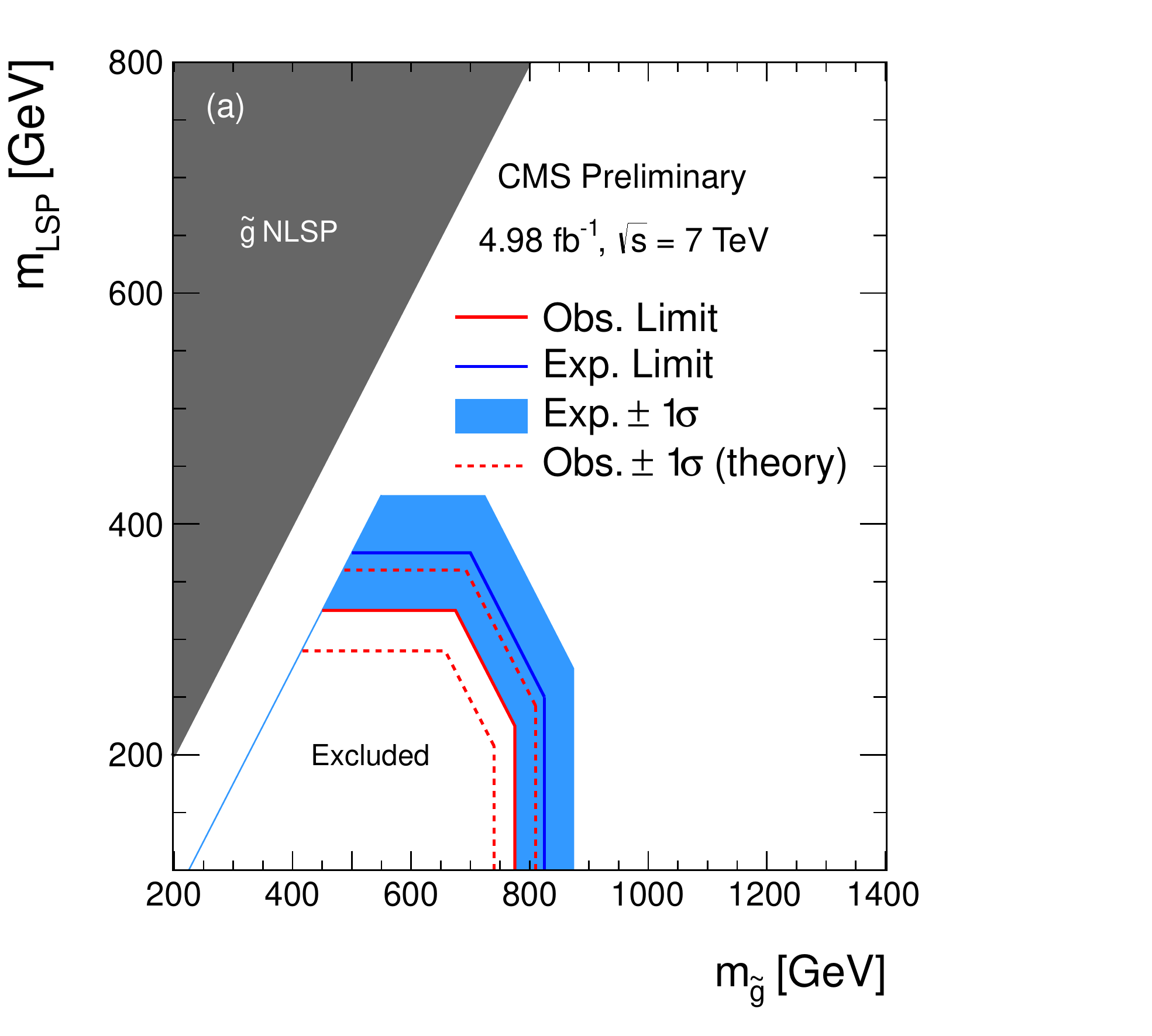}
\caption{The $95\%$ CL limit using the CLs technique for the CMSSM model with tan$\beta$=$40$, $A_{0}$=$0$ GeV and $\mu>0$ for the $\tau\tau$ final state (left) and cross section upper limits for the simplified model (right). The area below the solid lines are excluded. }
\label{fig:third}
\end{figure}

The final number of events selected in data is consistent with the predictions for SM processes, and no evidence of SUSY has been observed. Upper limits on the signal cross section with 95\% CL are set in the CMSSM model for a gaugino mass, $m_{1/2}$, of < 495 GeV is excluded at 95\% CL for scalar masses, $m_0$, < 400 GeV. This corresponds to a limit in the mass of the gluino of < 1.15 TeV. In the $\tau\tau$ final state, a gluino with mass < 740 GeV is excluded at 95\% CL for the simplified model, where gluinos are produced in pairs and subsequently decay to $\tau$-lepton pairs ($\tilde{g}\rightarrow q\bar{q}\tilde{\chi^{0}_{2}}$; $\tilde{\chi^{0}_{2}} \rightarrow \tau\bar{\tau} \tilde{\chi_{0}}$),~in Fig.~\ref{fig:third}. 

\section{Search for new physics in events with b-quark jets and missing transverse energy in pp collisions at $\sqrt{s}$= $7$ TeV}

The analysis is performed based on events with large missing transverse energy, at least three jets, and at least one, two, or three b-tagged jets. In these final states, the main source of SM  background are events with top quarks, comprising $t\bar{t}$ pair and single-top-quark events, events with a W or Z boson accompanied by jets, and multijet QCD events produced purely through strong-interaction processes. In order to evaluate the QCD background, the minimum azimuthal opening angle between the MET vector and each of the three highest-$p_T$ jets in an event has been used. Background from Z+jets events (Z $\rightarrow$ $\nu\bar{\nu}$) is evaluated by scaling the measured rates of Z $\rightarrow$ $l^{+}l^{-}$ (l=e, $\mu$) events. For $t\bar{t}$ and W+jets,  MET distributions have been determined from the different decay products of these processes.

\begin{figure}[hbt]
\centering
\includegraphics[height=6.5cm]{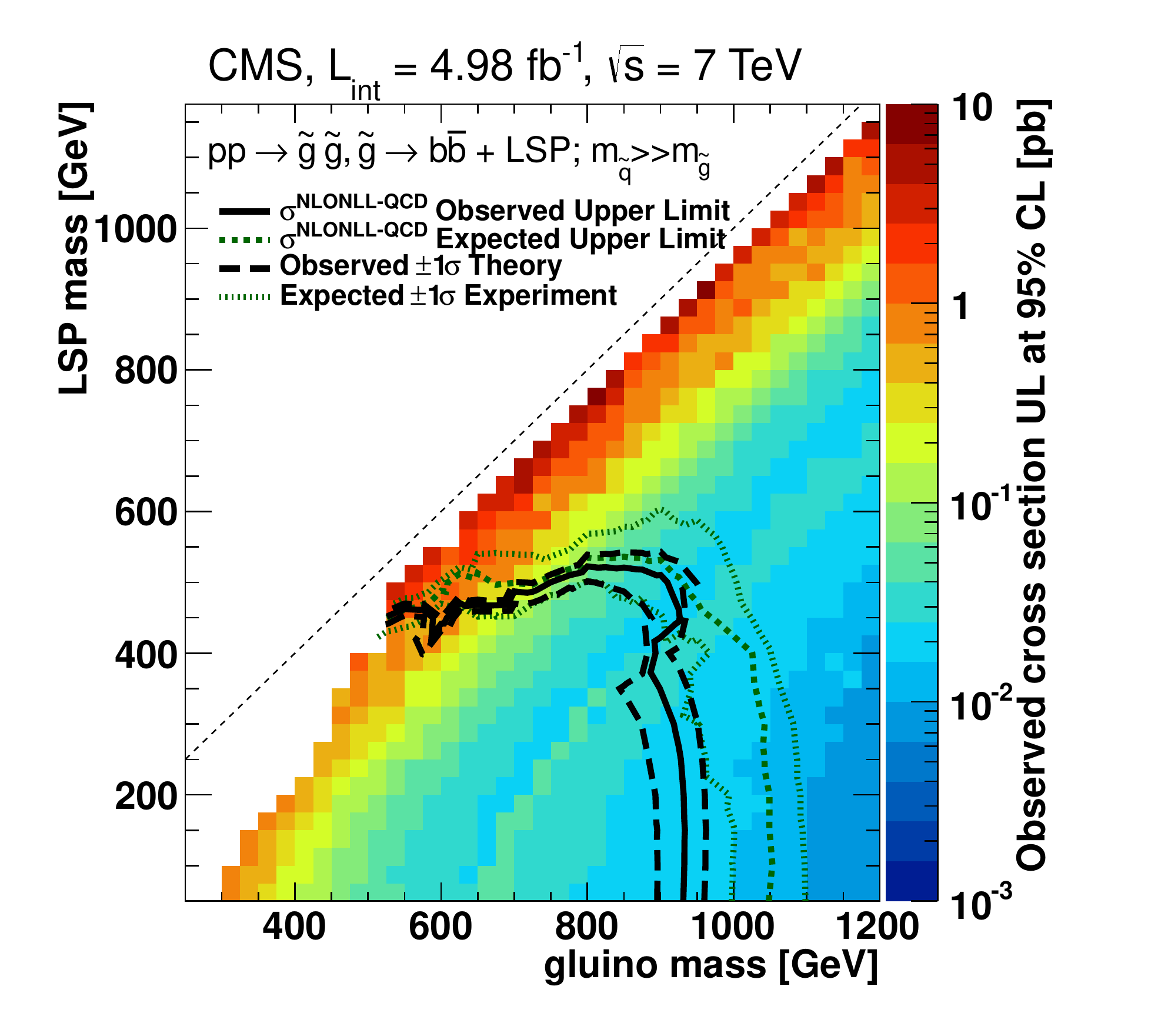}
\includegraphics[height=6.5cm]{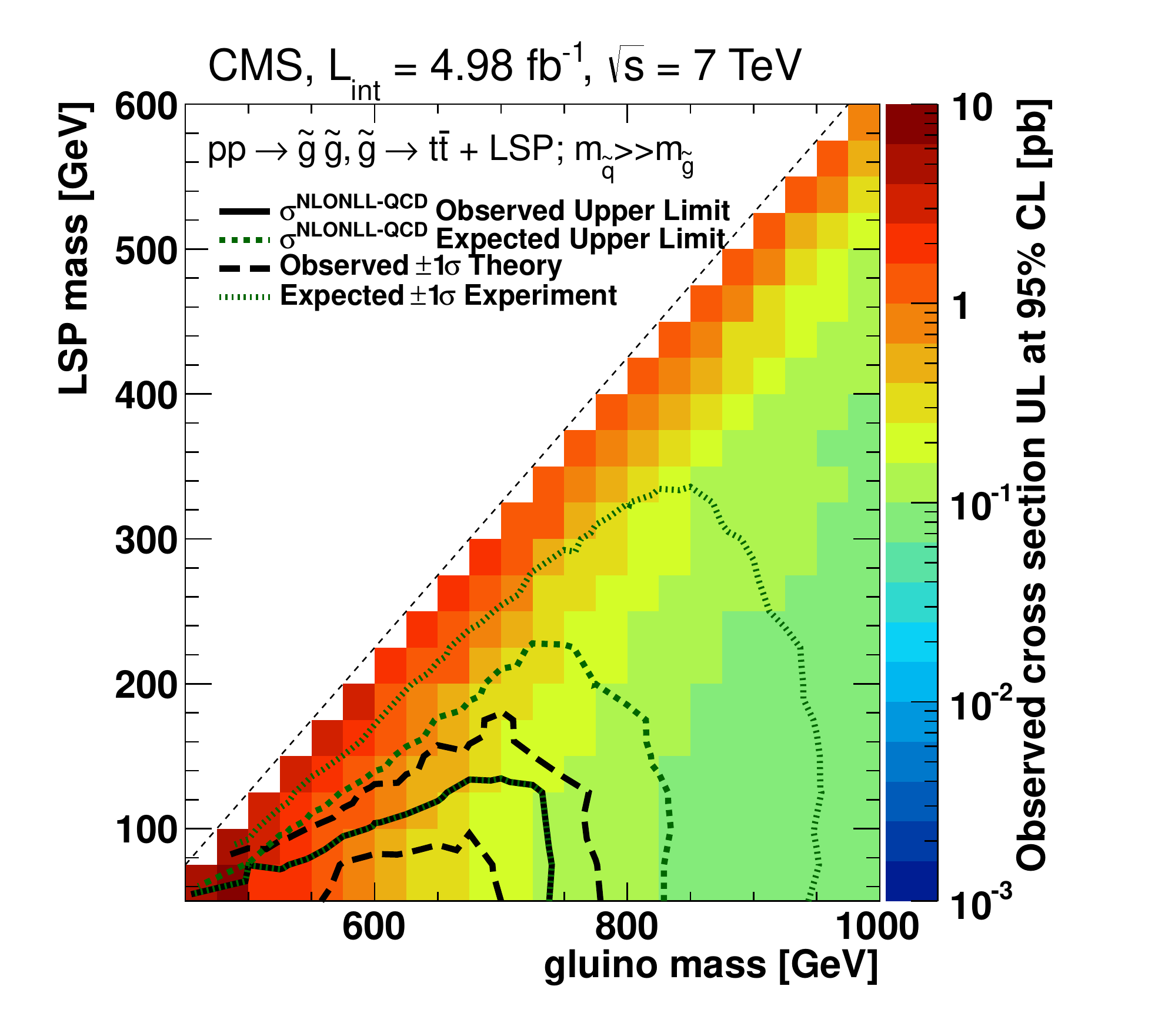}
\caption{The 95\% CL observed cross section upper limits for multi-bottom (left) and multi-top (right) simplified models. The kink in the curve (left) is caused by moving to a region where the observed limit is affected by the data being above the prediction in 2b-jets and 3b-jets search regions at the 95\% CL. Missing points along the diagonal are due to removal in cases where the uncertainty in ISR modeling appreciably affects the limit.}
\label{fig:fourth}
\end{figure}

No evidence for a significant excess of events beyond the expectations of the SM has been found in this channel. Limits on new physics in the context of the b-jet-rich simplified model spectra are shown in Fig.~\ref{fig:fourth} for multi bottom (left) and top (right) quark final states.

\section{Search for new physics in events with same-sign dilepton and b-tagged jets in pp collisions $\sqrt{s}$= $8$ TeV}
\begin{figure}[hbt]
\centering
\includegraphics[height=5cm]{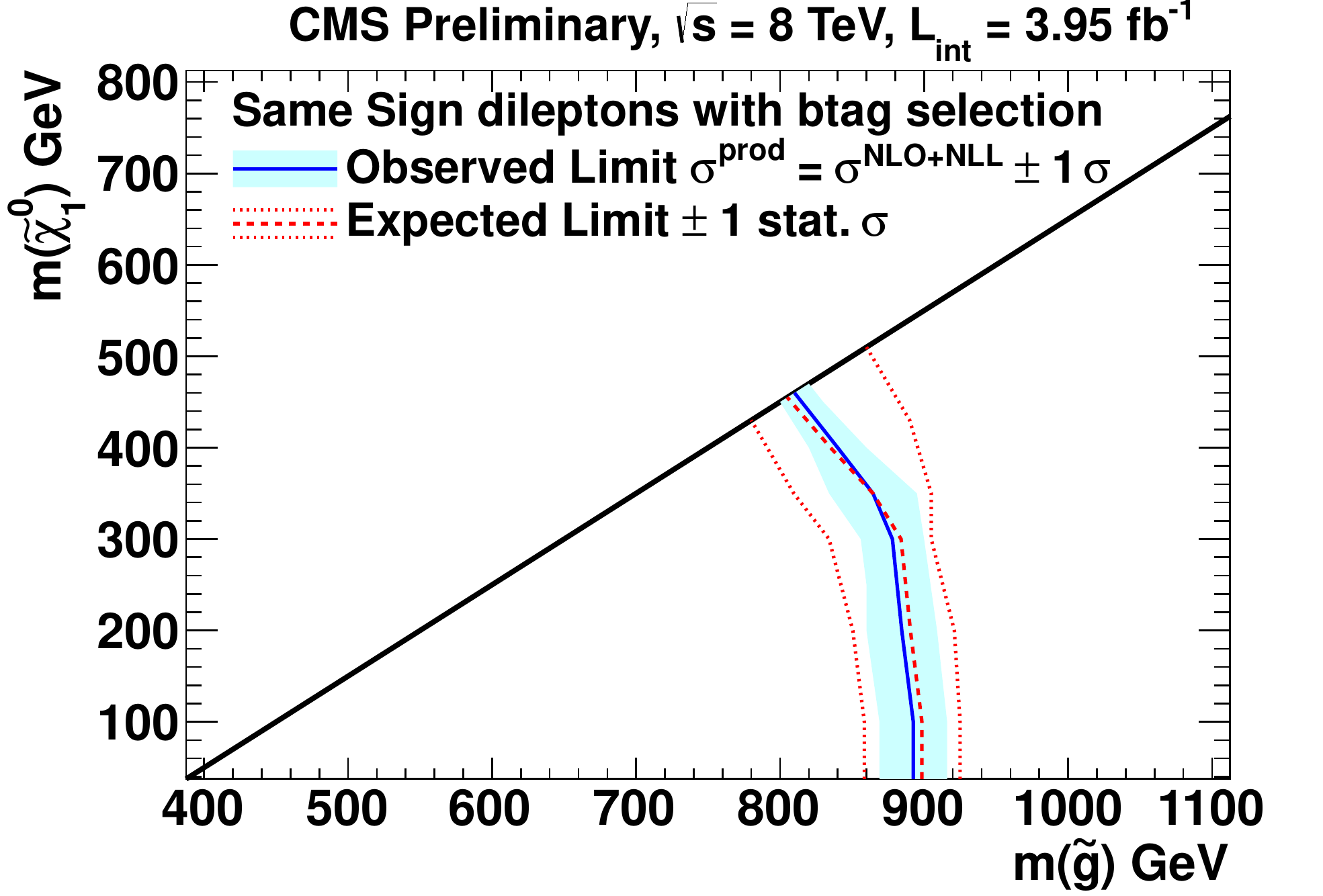}
\includegraphics[height=5cm]{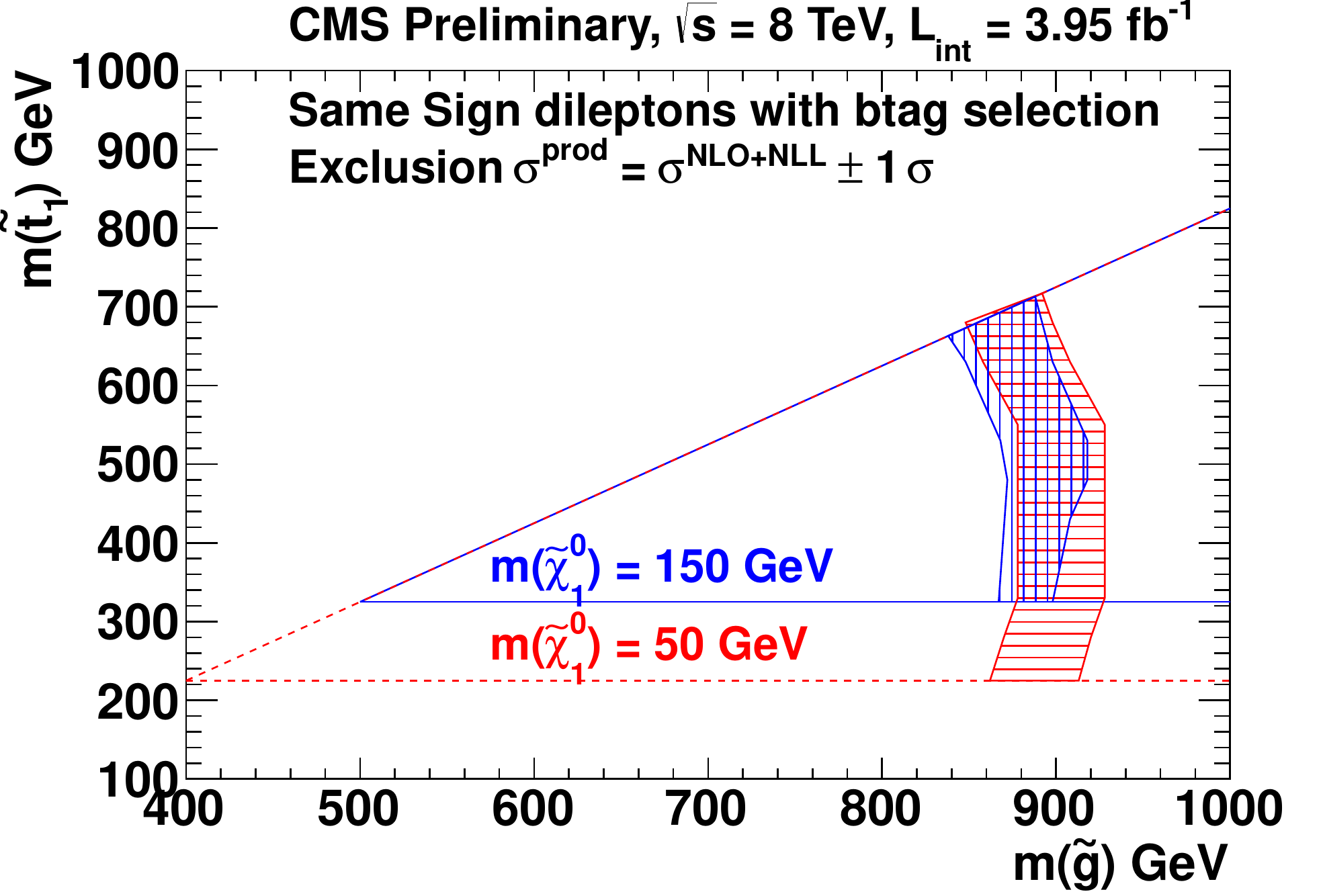}
\caption{The $95\%$ CL limit using the CLs technique for gluino decay via virtual stop quarks (left) and gluino decay to on-shell top squarks (right). The area below the solid lines are excluded.}
\label{fig:fifth}
\end{figure}

This analysis, a preliminary extension of the CMS same-sign dilepton search~\cite{sse}, is based on events with two isolated, high transverse momentum, same-sign leptons (e or $\mu$), and at least two b-tagged jets. The event counts in a number of signal regions defined in terms of additional requirements on MET, the number of b-tagged jets, and $H_{T}$, are compared with expectations from the SM. There are three primary sources of SM backgrounds considered in this signal region; fake leptons, which are mainly produced in heavy flavor decays, misidentified hadrons, muons from meson decay or electrons from unidentified conversions; events with opposite-sign isolated leptons where one of the lepton is an electron and its charge is misreconstructed due to severe bremsstrahlung in the tracker material; and rare SM processes that yield same-sign high $p_T$ leptons and b-jets. The background prediction method uses various $H_T$ and MET regions together with different trigger strategies. No evidence for an excess over the background prediction has been observed and the results have been used to set 95\% CL exclusion limits for the simplified models in Fig.\ref{fig:fifth}.

The results have been used to set upper limit on $\sigma$(pp $\rightarrow$ tt) < $0.39$ pb and  $\sigma$(pp $\rightarrow$ $\bar{t}\bar{t}$) < $1.51$ pb at 95\% CL. In addition, gluinos with masses up to approximately 880 GeV are excluded if they decay into stop or sbottom pairs.

\section{Summary}

Results of searches for SUSY in events with third-generation particles at the CMS experiment have been presented. The final number of events selected in data are consistent with the predictions for SM processes and no evidence of SUSY has been observed. The results of the hadronic and leptonic searches are interpreted in the context of the CMSSM and various simplified models.  In the absence of signal, limits on the allowed parameter space in the CMSSM and in simplified models were set, which exceed those set by previous analyses.

%The CMS Collaboration expects to collect ~L=$25$ $fb^{-1}$ of data by the end of 2012.

\end{document}